\documentclass[aps,prl,superscriptaddress,showpacs,twocolumn]{revtex4-1}
\usepackage[utf8]{inputenc}
\usepackage{graphicx}
\usepackage{amsmath}
\usepackage{amssymb}
\usepackage{xspace}
\usepackage{bm}
\usepackage[colorlinks,linkcolor=blue,citecolor=blue,urlcolor=blue]{hyperref}
\usepackage{braket}

\newcommand{\blue}[1]{{\color{black} #1}}

\begin{document}

\title{On-chip III-V monolithic integration of heralded single photon sources \\ and beamsplitters}

\author{J. Belhassen}
\affiliation{Laboratoire Matériaux et Phénomènes Quantiques, Université Paris Diderot, CNRS-UMR 7162, Paris 75013, France}
\author{F. Baboux}
\thanks{florent.baboux@univ-paris-diderot.fr}
\affiliation{Laboratoire Matériaux et Phénomènes Quantiques, Université Paris Diderot, CNRS-UMR 7162, Paris 75013, France}
\author{Q. Yao}
\affiliation{Laboratoire Matériaux et Phénomènes Quantiques, Université Paris Diderot, CNRS-UMR 7162, Paris 75013, France}
\author{M. Amanti}
\affiliation{Laboratoire Matériaux et Phénomènes Quantiques, Université Paris Diderot, CNRS-UMR 7162, Paris 75013, France}
\author{I. Favero}
\affiliation{Laboratoire Matériaux et Phénomènes Quantiques, Université Paris Diderot, CNRS-UMR 7162, Paris 75013, France}
\author{A. Lemaître}
\affiliation{Centre de Nanosciences et de Nanotechnologies, CNRS, Univ. Paris-Sud, Université Paris-Saclay, C2N–Marcoussis, F-91460 Marcoussis, France}
\author{W.S. Kolthammer}
\affiliation{Clarendon Laboratory, University of Oxford, Parks Road, Oxford OX1 3PU, United Kingdom}
\author{I.A. Walmsley}
\affiliation{Clarendon Laboratory, University of Oxford, Parks Road, Oxford OX1 3PU, United Kingdom}
\author{S.~Ducci}
\affiliation{Laboratoire Matériaux et Phénomènes Quantiques, Université Paris Diderot, CNRS-UMR 7162, Paris 75013, France}

\begin{abstract}

We demonstrate a monolithic III-V photonic circuit combining a heralded single photon source with a beamsplitter, at room temperature and telecom wavelength. Pulsed parametric down-conversion in an AlGaAs waveguide generates counterpropagating photons, one of which is used to herald the injection of its twin into the beamsplitter. 
We use this configuration to implement an integrated Hanbury-Brown and Twiss experiment, yielding a heralded second-order correlation $g^{(2)}_{\rm her}(0)=0.10 \pm 0.02$ that confirms single-photon operation. The demonstrated generation and manipulation of quantum states on a single III-V semiconductor chip opens promising avenues towards real-world applications in quantum information.

\end{abstract}

\maketitle

Integrated photonic circuits provide a promising approach to achieving a wide range of quantum information tasks. Compared to free-space optics, chip-based photonics offers crucial advantages in terms of portability, stability and scalability. In particular, photonic chips lie at the heart of the linear optical quantum computing scheme \cite{Knill01}, which provides a toolbox to realize all-optical processing tasks using solely single photon sources and detectors, and elementary linear components such as beamsplitters and phase shifters.

In this context, rapid progress has been made in recent years to develop integrated circuits that achieve on-chip quantum interference, entanglement and gate operations \cite{OBrien09,Politi08,Smith09,Peruzzo11,Tillmann13,Spring13,Crespi13}. However, such demonstrations usually rely on external sources to generate quantum states of light, which are then fed into passive circuitry. On the other hand, great efforts have been made to develop miniaturized sources of single photons. 
While single-emitter systems \cite{Aharonovich16}, such as quantum dots have made remarkable progress as bright and deterministic single-photon sources \cite{Somaschi16}, parametric non-linear processes offer an unmatched flexibility in wavelength and bandwidth, as well as a capability of constructing many identical sources. The latter devices operate in a heralded configuration, in which pairs of twin photons are generated and the detection of one photon is used to herald the existence of the other \cite{Hong86}.
Such heralded single photon sources have been realized in an integrated manner, e.g. exploiting parametric down-conversion in PPKTP \cite{URen04} or PPLN \cite{Pomarico12,Ngah15} waveguides, or four-wave mixing in silicon \cite{Sharping06,Takesue07} and silica \cite{Spring13b,Spring17} waveguides, to produce single-photon states of high purity and indistinguishability, at room temperature and telecom wavelength. Several parametric sources can be integrated on a single chip, paving the way to large scale operations \cite{Meany14,Silverstone14,Vergyris16,Spring17}.

An important challenge now is to combine the progress on sophisticated linear circuits and high-performance single-photon sources to achieve generation and manipulation on a single chip. First experimental results in this direction are promising, either with hybrid technology -- e.g. PPLN sources with laser-written glass circuits \cite{Meany14,Vergyris16} or III-V quantum dots with silica circuits \cite{Murray15} --  or with monolithic technology on PPLN \cite{Krapick13,Kruse13,Solntsev14,Kruse15}, Silicium \cite{Silverstone14,Spring17} or GaAs \cite{Prtljaga14,Jons15,Rengstl15,Bentham15,Schwartz16,Prtljaga16}.

In particular, the recent integration of III-V quantum dots and beamsplitters  on GaAs points to a notably flexible platform \cite{Prtljaga14,Jons15,Rengstl15,Bentham15,Schwartz16,Prtljaga16}. Indeed, GaAs offers several assets for integrated quantum photonics \cite{Dietrich16,Orieux17}: it can host both two-level \cite{Yuan02,Somaschi16} and parametric emitters \cite{Lanco06,Horn12,Horn13}, its direct band gap enables electrically injected sources \cite{Yuan02,Boitier14}, its high electro-optic effect allows for fast phase shifters \cite{Wang14}, and it can be combined with superconducting nanowires to achieve on-chip detection \cite{Sprengers11}.

Here, we report the first realization of a monolithic GaAs/AlGaAs photonic circuit combining a parametric heralded single-photon source with a beamsplitter. This configuration allows us to realize an integrated Hanbury-Brown and Twiss experiment -- with on-chip beam splitting and off-chip detection -- which confirms single-photon generation and manipulation. This realization, at room temperature and telecom wavelength, highlights the potential of the GaAs platform to realize fully integrated quantum circuits. 
The manipulation of quantum states, here demonstrated for single photons, could in principle be extended to the rich variety of entangled states commonly produced by parametric waveguide sources \cite{Horn13,Orieux13,Autebert16,Kang16}, paving the way to more complex applications.

\begin{figure}[t]
\centering
\includegraphics[width=1\columnwidth]{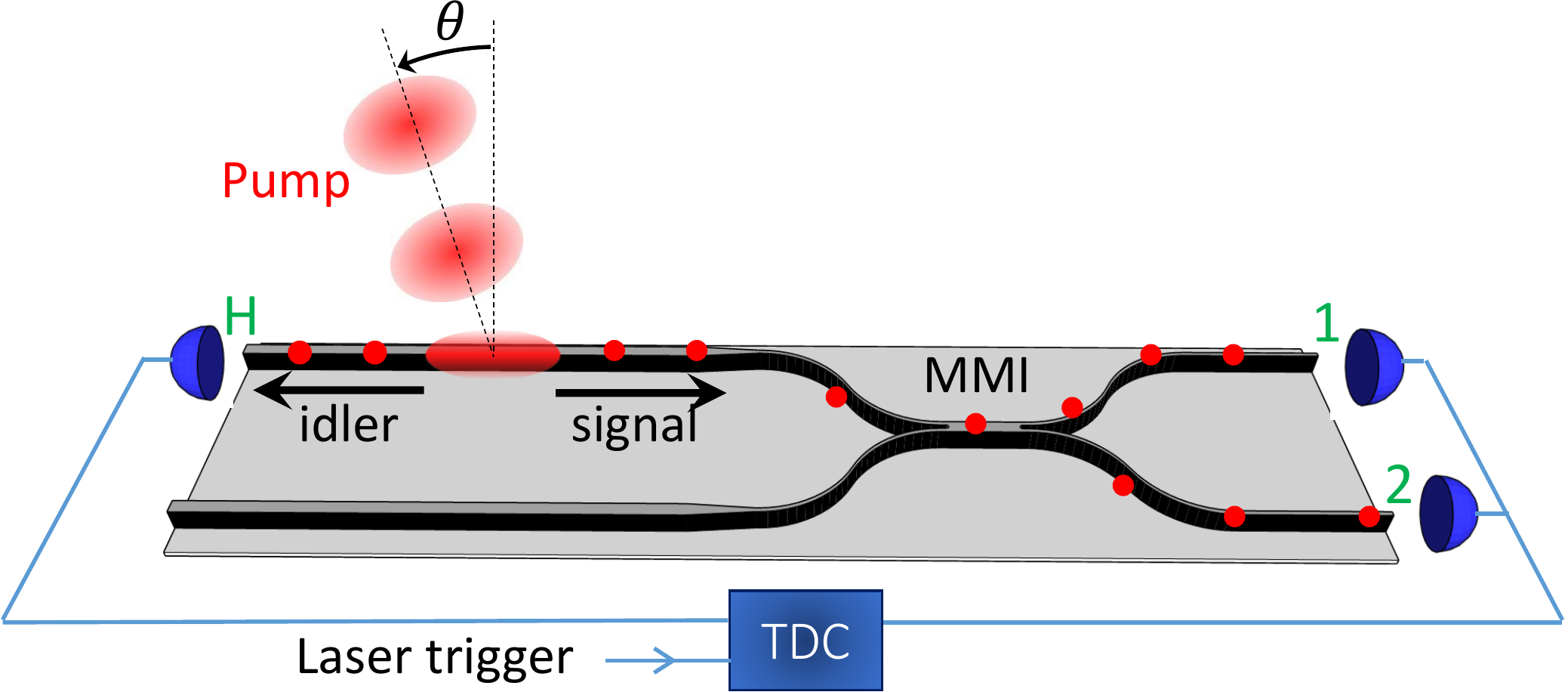}
\caption{
Working principle of the device. A transverse pump laser impinges on the input waveguide and generates pairs of counterpropagating photons. One photon exits port H and is used to herald the injection of its twin into the multi-mode interferometer (MMI) coupler. Detectors placed at ports H, 1, and 2 measure photon correlations using a time-to-digital converter (TDC).
}
\label{Fig1}
\end{figure}

The working principle of the device is sketched in Fig.~\ref{Fig1}. The device is made of four waveguides that meet in a central 2x2 coupler.
Photon pairs are generated through spontaneous parametric down-conversion (SPDC) by pumping one of the input waveguides (labelled H) with a pulsed laser beam impinging from the top, making an angle $\theta$ with the vertical.
A pair of counterpropagating, orthogonally polarized telecom-band photons are generated by SPDC in a type-II counterpropagating phase-matched process \cite{DeRossi02,Orieux13}. As usual, the wavelengths of the generated photons are determined by energy and momentum conservation. Of the two non-linear interactions occurring in the waveguide \cite{DeRossi02,Orieux13}, in our work we consider the one that generates a TE-polarized idler photon propagating away from the coupler and a TM-polarized signal photon propagating towards it.
The idler photon is detected with an avalanche photodiode (APD) connected to port H, which heralds injection of the signal photon into the beamsplitter, consisting in a multi-mode interfermometer (MMI).

\begin{figure}[t]
\centering
\includegraphics[width=0.8\columnwidth]{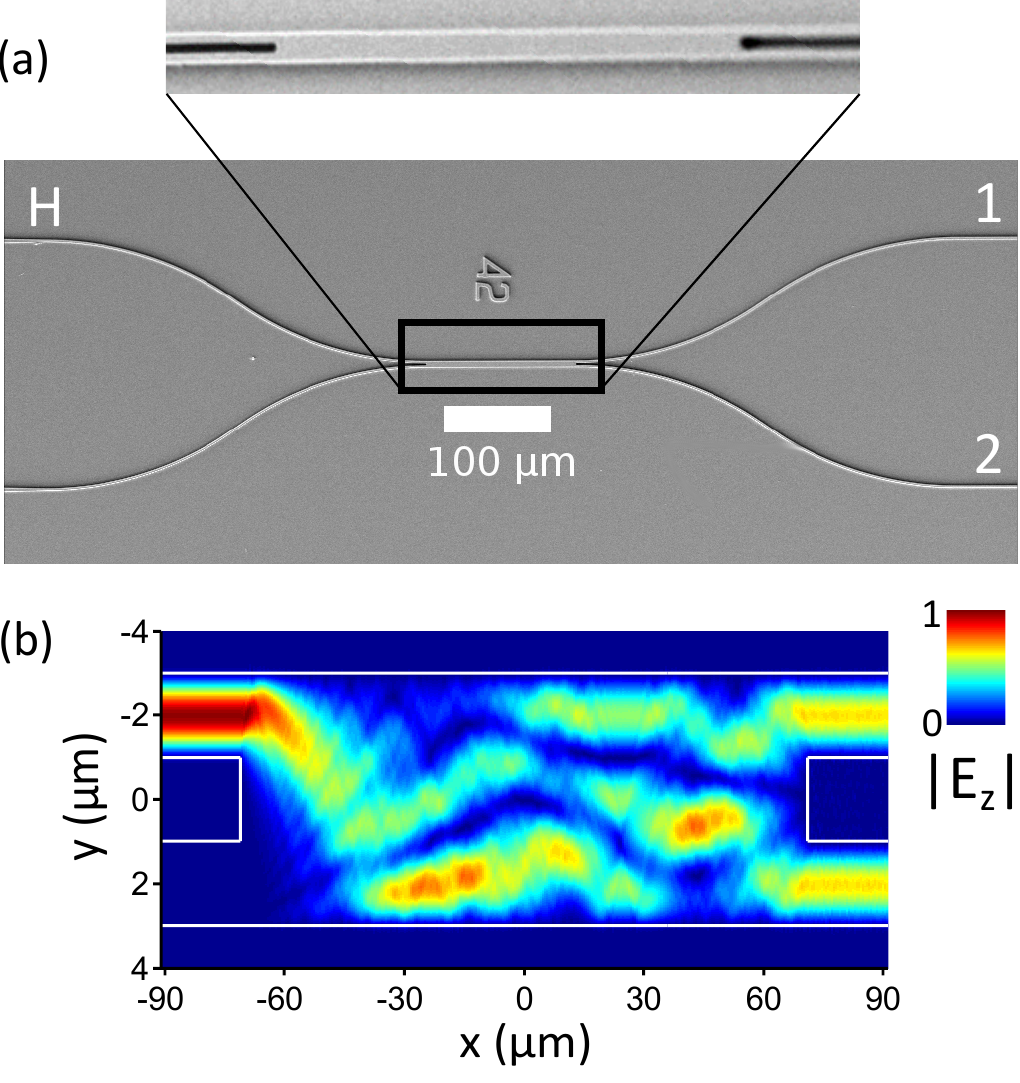}
\caption{
(a) Scanning electron microscope image of the fabricated device, with a close-up of the beamsplitter region.
(b) Simulation of the \blue{z-component of the electric field (absolute value) inside the multi-mode interferometer when injecting the fundamental TM mode into waveguide H.}
}
\label{Fig2}
\end{figure}

The AlGaAs device was fabricated by e-beam lithography followed by dry etching. 
Fig. \ref{Fig2}a shows a scanning electron microscope image of the central part of the sample. The total length of the device is $3$ mm, and the input and output waveguides are separated by 250 $\mu$m. The input waveguides (on the left) have a 6-$\mu$m width in their straight portion, designed to optimize the non-linear conversion. This width is then adiabatically decreased to 2 $\mu$m to ensure guiding of only a single spatial mode. An S-bend with a radius of 300 $\mu$m then guides the photons to the MMI (close-up in Fig. \ref{Fig2}a). The two output waveguides have a 2 $\mu$m-width.

The epitaxial structure of the device (see Supplementary Material) consists of alternating AlGaAs layers with different Al concentrations. 
The core implements a vertical quasi-phasematching for the pump beam. It is surrounded by two distributed Bragg reflectors, that act as a cladding for the telecom photons, and also define a vertical cavity that confines the pump beam to enhance the conversion efficiency \cite{Orieux11}.

The MMI coupler is designed to split the input signal field equally between the two output waveguides. \blue{A simulation of the electric field in the structure, using the beam propagation method, is shown in Fig. \ref{Fig2}b. In the upper left, the fundamental TM mode injected into the upper input waveguide} enters the $6$-$\mu$m wide central region of the MMI, where it decomposes into a superposition of multiple modes with differing propagation constants. The interference between these modes leads to a splitting of the optical field between the two outputs. The length of the MMI ($142 \,\mu$m) is designed to reach a balanced splitting ratio. \blue{Compared to other beamsplitter geometries such as evanescent couplers, the MMI geometry offers a better tolerance to polarization and wavelength variations \cite{Soldano95,Rajarajan99}}; it can also be easily generalized to a higher number of intput and output waveguides for more complex applications \cite{Peruzzo11}.

\begin{figure}[t]
\centering
\includegraphics[width=0.85\columnwidth]{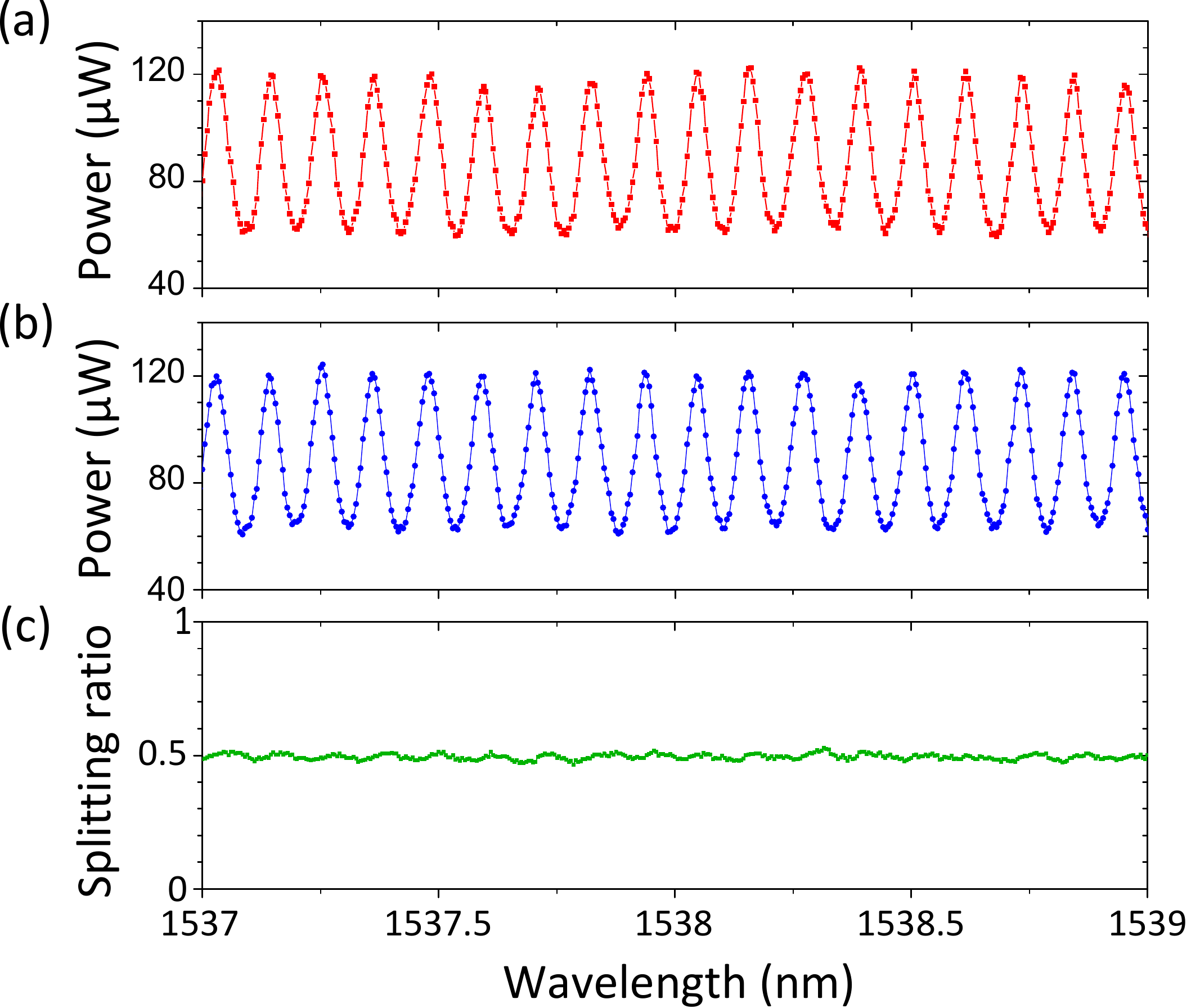}
\caption{
(a) Power transmitted through the output waveguide 1, as a function of the wavelength, when injecting TM-polarized laser light in port H of the device (see Fig. 1). 
(b) Same measurement carried out for the output waveguide 2.
(c) Deduced splitting ratio of the beamsplitter in TM polarization.
}
\label{Fig3}
\end{figure}

We start with a classical characterization of the device to evaluate its optical losses and splitting ratio.
We inject TM-polarized telecom laser light (to simulate the signal field) into port H, and we monitor the transmission through the output ports 1 and 2 as the laser wavelength is varied.
The sample is thermally stabilized at $T=22^{\circ}$, and the injection and collection is done using x40 microscope objectives.
Fig. \ref{Fig3}a and b shows the power transmitted through outputs 1 and 2, respectively. The observed oscillations in the transmitted power correspond to a Fabry-Perot cavity effect due to the facets reflectivity \cite{Orieux11}.
The contrast of these interference fringes allow determining the propagation loss coefficient $\alpha$ without requiring knowledge of the coupling efficiencies \cite{DeRossi05}. We extract $\alpha \simeq 1.3 $ cm$^{-1}$ consistently from the data of Fig. \ref{Fig3}a and \ref{Fig3}b.
This value accounts for all propagation losses within the device, including the input waveguide, MMI region, and output waveguides, and it compares very favourably with recently reported GaAs active photonic circuits \cite{Rengstl15,Jons15,Schwartz16}.
Finally, the  measurements of Figs. \ref{Fig3}a-b together allow us to determine the beamsplitter splitting ratio \blue{by dividing the power transmitted through port 1 by the sum of the power transmitted through ports 1 and 2. We obtain a ratio of $(49.5 \pm 0.9) \%$, in agreement with the simulations of Fig. \ref{Fig2}b.}
Note that the device could also be operated in TE polarization, for which we measured $\alpha \simeq 0.9 $ cm$^{-1}$ and a splitting ratio of  $(49.2 \pm 1.0) \%$.

We now proceed to the quantum characterization of the device. In all following measurements, the device is pumped with a pulsed 769 nm laser to generate photon pairs through SPDC (see Fig. \ref{Fig1}). 
We first consider possible correlations between the down-converted photon pairs. In general, energy and momentum conservation in SPDC can lead to entanglement into the spatial and spectral modes of the photons. However if detection of the heralding photons is not mode-resolved, the heralded signal photons are left into a mixed state, hence reducing the purity and indistinguishability of the heralded source \cite{Grice01,Mosley08}. 
This issue is usually circumvented by using spatial and spectral filters, but at the price of reducing the brightness. By contrast, in our device photon pairs can be directly produced in a nearly separable state. Indeed, while the guided geometry ensures spatially monomode emission, the spectral correlations can be precisely tailored via the pump beam properties (pulse duration and spot size) \cite{Caillet09,Boucher15}. Given the pulse duration of our Ti:Sa laser (5 ps), we estimate that a  frequency-separable state should be obtained for a spot size of 1.5 mm (intensity FWHM along the waveguide direction).

\begin{figure}[t]
\centering
\includegraphics[width=0.95\columnwidth]{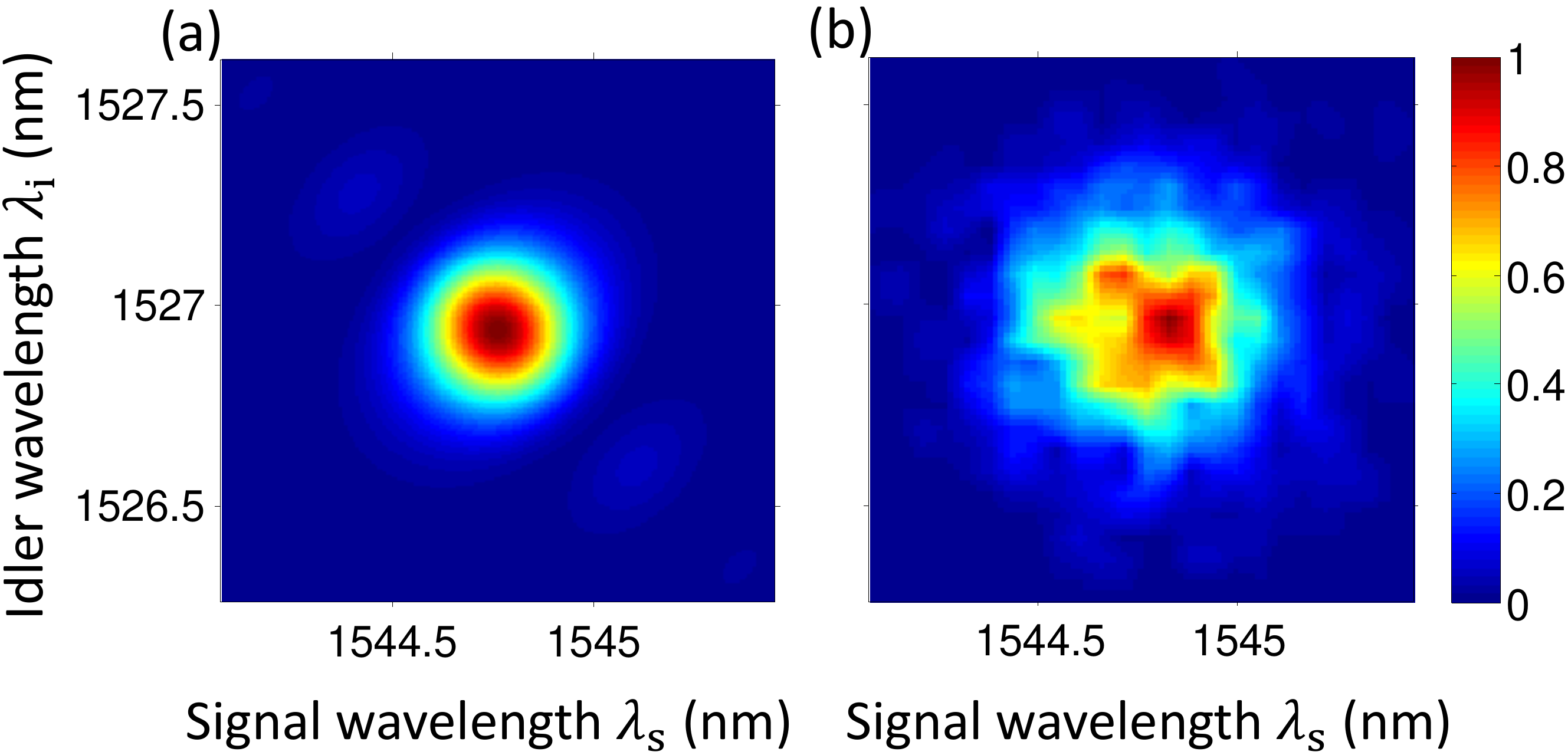}
\caption{
(a) Calculated and (b) measured joint spectral intensity of the SPDC source, evidencing the emission of nearly frequency-uncorrelated photons. \blue{The value of the pump incidence angle $\theta \simeq 1.4^{\circ}$ can be deduced from the measured signal and idler central wavelengths \cite{Caillet09,Orieux13}.}
}
\label{Fig4}
\end{figure}

Figure \ref{Fig4}a shows the joint spectral intensity (JSI) calculated \cite{Boucher15} for these pump parameters and our device characteristics. For simplicity we neglected the Fabry-Perot effect due to the facets reflectivity. The shape is close to the circular shape characteristic of a frequency-separable state. More quantitatively, we can extract the Schmidt number $K$, which measures the effective number of orthogonal frequency modes spanned by the biphoton wavefunction: the simulation yields $K = 1.2$, close to the $K=1$ value of a perfectly separable state.

To verify these predictions, we measure the joint spectral intensity (JSI) of the biphoton by using two single-photon spectrometers \cite{Avenhaus09,Eckstein14}. The upper input waveguide (see Fig. \ref{Fig1}) is pumped using an elliptical spot of length 1.5 mm (along the waveguide direction), pulse duration of $5$ ps and repetition rate of $3.8$ MHz.
Idler photons are collected through waveguide H, and signal photons can be collected indifferently through waveguides 1 or 2: here we collect them through waveguide 1.
Two polarizers are used to select only the nonlinear interaction that produces TE-polarized idler and TM-polarized signal photons. A long-pass filter is used at port H to reject luminescence induced by the pump beam.
Each photon is sent to a time-of-flight fiber spectrograph built from a spool of optical fiber with high chromatic dispersion and an APD (free-running IDQ 220) \cite{Avenhaus09,Eckstein14}.
A time-to-digital converter records their arrival time ($t_s$ and $t_i$ for signal and idler) and the trigger signal received from the laser pulse picker 
($t_{\rm laser}$). 
The time delays $\tau_{s/i}=t_{s/i}-t_{\rm laser}$ are then converted into photon wavelengths $\lambda_{s/i}$. 
This procedure yields the experimental joint spectral intensity shown in Fig. \ref{Fig4}b.
A circular shape is observed, slightly broadened by the spectral resolution of the spectrograph ($0.2$ nm).
From this data a lower bound $K > 1.05$ is obtained for the Schmidt number, compatible with the simulated $K=1.2$.

Second-order correlation function measurements are then performed to complete the quantum characterization of the device. Since our APD resolution time (250 ps) is greater than the duration of the signal and idler wavepackets ($\simeq 5 $ ps, inferred from the JSI measurement), time-integrated correlation functions can be obtained, which provide a different route to infer the spectral correlations of emitted photon pairs \cite{Christ11}.
In the following, since the spectral width of the biphotons is of the order of 0.5 nm (see Fig. \ref{Fig4}a), we use 1.2 nm interferometric filters, instead of polarizers, to select the desired interaction: this allows to filter out noise related to incoherent emission (e.g. substrate luminescence) while leaving the biphoton spectrum unaffected.

We first determine the coincidence-to-accidental ratio (CAR) by measuring the cross-correlation function $g_{s,i}^{(2)}(0)=1+\text{CAR}$ between the signal and idler fields.
For this we measure the temporal correlations between ports H and 1 of the device (see Fig. \ref{Fig1}) during a 1h acquisition time. 
The cross-correlation function is proportional to the measured coincidences, $g_{s,i}^{(2)}(0)=\gamma \, N_{1H}$, and the normalization factor $\gamma$ is determined by measuring the correlations at long time delays, where the cross-correlation function reaches unity.
For a mean incident pump power of 60 mW (measured before the injection lens), we obtain $g_{s,i}^{(2)}(0)= 44 \pm 3$. Here and in the following, detector counts have been integrated within a 2.5 ns acceptance window around the mean photon arrival delays, and corrected for background noise. From the measured cross-correlation function we can extract \cite{Christ11} the mean photon number $\langle n \rangle \simeq 1/g_{s,i}^{(2)}(0)= 0.023 \pm 0.002$. This confirms that we are in a low pump regime where multipair emission is strongly reduced, as required to obtain an heralded single-photon source of high photon-number purity.

We next investigate the auto-correlation function $g_{s,s}^{(2)}(0)$ of the signal field by measuring the temporal correlations between ports 1 and 2 of the device. This corresponds to a Hanbury-Brown and Twiss measurement with a chip-integrated light source and beamsplitter. 
The measurement yields $g_{s,s}^{(2)}(0) =1.7 \pm 0.3$. This auto-correlation function yields an additional determination \cite{Christ11} of the Schmidt number $K$, since $g_{s,s}^{(2)}(0)\simeq 1+1/K$. We deduce  $K=1.4 \pm 0.3$,
in agreement with the nearly single-mode emission suggested by the measured and calculated joint spectrum.

We now turn to the  measurement of the heralded auto-correlation function \cite{Hong86,Grangier86}. For this we use an additional APD (gated IDQ 210) at port H, which is triggered by the laser pulse picker, and we perform correlation measurements between ports H, 1 and 2 (see Fig. \ref{Fig1}).
\blue{Integrating during a 6h acquisition time the number of heralding events ($N_H=17 \, 128 \,410$), two-fold coincidences ($N_{1H}=49 \,729$ and $N_{2H}=88 \, 262$) and three-fold coincidences ($N_{12H}=25$),} we obtain \cite{Grangier86} $g^{(2)}_{\rm her}(0)=(N_{12H} N_H)/(N_{1H} N_{2H})=0.10 \pm 0.02$. 
This result can be compared to a theoretical estimate based on the unheralded measurements, $g_{\rm her, th}^{(2)}(0) \simeq 2 \, \langle n \rangle \, g_{s,s}^{(2)}(0)$ (see Supplementary Material). This yields $g_{\rm her,th}^{(2)}(0)=0.08 \pm 0.02$, in good agreement with the direct measurement of the heralded auto-correlation. These results show that the heralded source emits single photons with high modal purity, a minimal multi-photon component, and that the single-photon character is preserved during the guiding and splitting process within the device.
Taking into account the collection efficiency (from chip to APD) and APD detection efficiency (15 and 20 $\%$ respectively), the emission rate of single-photons at the output of the chip is  $\simeq 200$ Hz for this pump power. \blue{In the future, this rate could be improved by several means: using the full repetition rate of the pump laser (76 MHz), depositing anti-reflection coating on the end facets (to reach near-unity transmission), and using a superconducting detector (with typical efficiency 70 \%) to enhance the heralding efficiency would allow reaching a 20 kHz single photon emission rate. On the other hand, using lower instantaneous pump powers} should reduce the $g^{(2)}_{\rm her}(0)$ value, by further suppressing multipair emission. Measurements of the cross-correlation function as a function of pump power $P$ (see Supplementary Material) indicate an essentially linear increase of $\langle n \rangle$ with $P$, corroborating the expectation of increased single-photon purity at smaller pump power, which could be useful for specifically demanding applications.

In summary, we have reported the first demonstration of a monolithic GaAs/AlGaAs photonic circuit combining a parametric heralded single-photon source and a beamsplitter.
This device allows performing an integrated Hanbury-Brown and Twiss experiment that confirms single-photon generation and manipulation within the same circuit, at room temperature and telecom wavelength. The used transverse pump configuration circumvents the usual issue of pump filtering (required in collinear injection schemes), allows a direct spatial separation of the heralding and heralded photons, and a tuning of the joint spectral intensity to obtain a nearly separable state ($K \sim 1$) ensuring high single-photon purity.
The demonstrated scheme can be extended to more complex photonic operations: for instance, pumping both input waveguides of the device would enable an integrated, heralded Hong-Ou-Mandel experiment, to test the indistinguishability between remote parametric sources, a key requirement for scalability \cite{Knill01,Spring17}. \blue{The device integration could be pushed further by using an electrically pumped rectangular VCSEL \cite{Gronenborn11} on top of the circuit to pump the parametric generation.} Finally, depositing electrodes should enable the implementation of fast phase shifters \cite{Wang14} to manipulate the produced quantum states and progress further towards optical quantum computing tasks on the GaAs platform.

\bigskip

\blue{See \textbf{Supplementary Material} for details on the sample structure, additional experimental data and derivation of analytical formulas for the autocorrelation function.}

\bigskip

\textbf{Acknowledgments}: We thank G. Boucher and Y.~Halioua for their contribution to the early stage of this project. 
We acknowledge support from Délégation Générale de l’Armement (funding of J. Belhassen), Agence Nationale de la Recherche (project SEMIQUANTROOM), Région Ile-de-France
in the framework of the C’Nano DIM NanoK (project SPATIAL), the Fondation Franco-Chinoise pour les Sciences et Applications (FFCSA), the European Research Council (ERC) through the Ganoms project (No.306664) and the French RENATECH network. I.A.W. acknowledges support from the EU via an ERC Advanced Grant (MOQUACINO). I.A.W and W.S.K. were funded also by the UK Engineering
and Physical Sciences Research Council through  Programme Grant No. EP/K034480/1, and the EPSRC NQIT Quantum Technology Hub.

\bibliographystyle{Science}
\bibliography{Biblio}

\clearpage
\onecolumngrid
\begin{center}
\textbf{\large Supplemental Material}
\end{center}

\setcounter{equation}{0}
\setcounter{figure}{0}
\setcounter{table}{0}
\makeatletter
\renewcommand{\theequation}{S\arabic{equation}}
\renewcommand{\thefigure}{S\arabic{figure}}

\section{Epitaxial structure of the sample}

\begin{figure}[h]
\centering
\includegraphics[width=0.3\columnwidth]{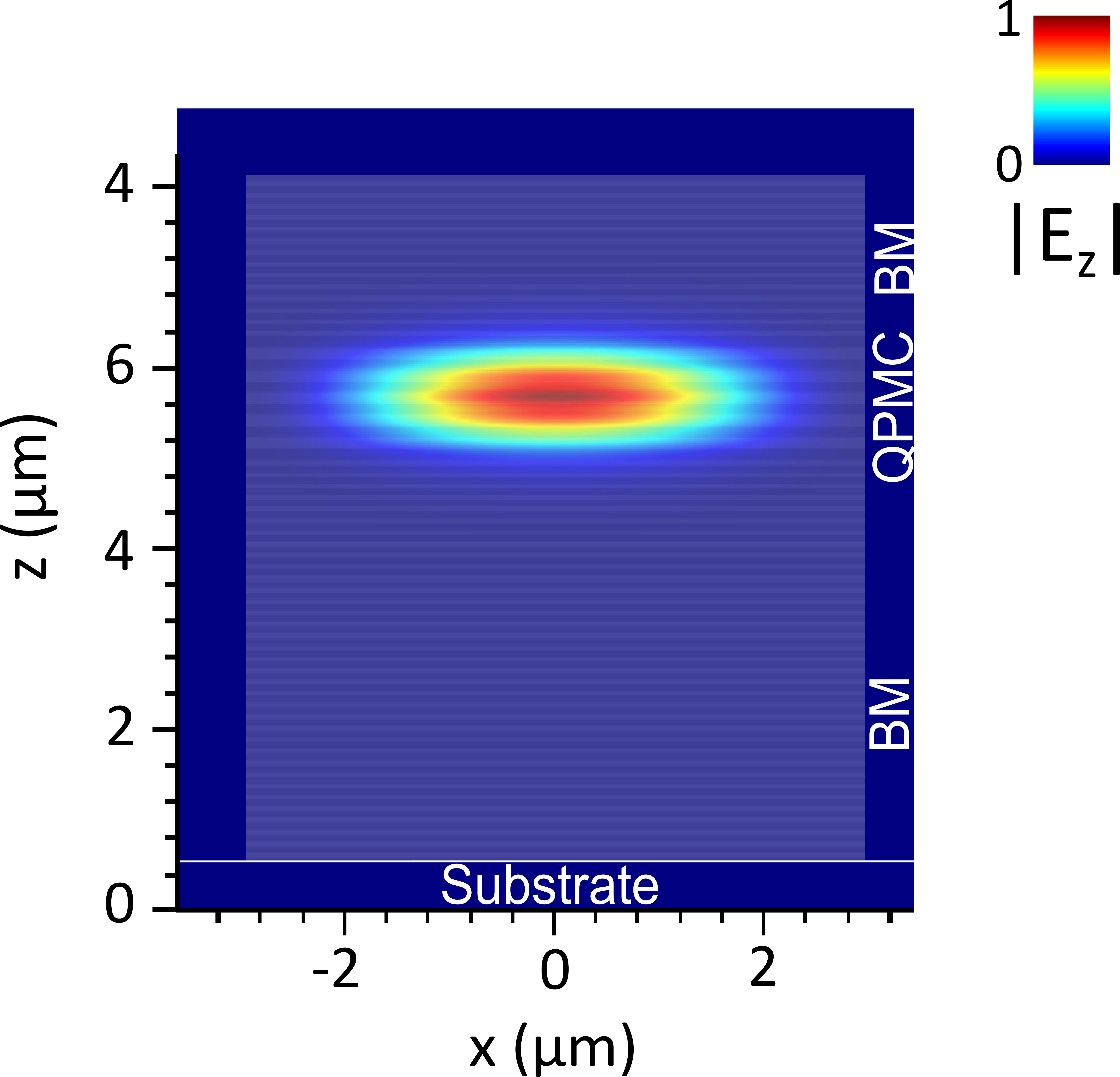}
\caption{
Simulated electric field profile (absolute value of the z-component) of the fundamental TM mode of the input waveguide, superimposed with the nominal epitaxial structure.
}
\label{Fig_structure}
\end{figure}

The epitaxial structure of the sample (see Fig. \ref{Fig_structure}) consists of a quasi-phasematching core, made of a 4.5-period Al$_{0.80}$Ga$_{0.20}$As/Al$_{0.25}$Ga$_{0.75}$As stacking, and two distributed Bragg reflectors, made of 36- and 14-period Al$_{0.90}$Ga$_{0.10}$As/Al$_{0.35}$Ga$_{0.65}$As stacking for the bottom and top mirrors, respectively. The distributed Bragg reflectors provide both a cladding for the down-converted telecom photons, and a vertical microcavity for the pump field, yielding a typical non-linear conversion efficiency of $10^{-11}$ pairs/pump photon for a mm-long sample \cite{Orieux11}.

Figure \ref{Fig_structure} shows a simulation of the TM$_{00}$ guided mode corresponding to the signal photons ($\lambda \sim \! 1.54 \,\mu$m) in the input waveguide of the device, superimposed with the nominal epitaxial structure.

\section{Impact of fabrication imperfections on the device linear properties}

\blue{We note that the MMI geometry introduces a $\pi/2$ phase difference \cite{Peruzzo11} between the two outputs after one passage through the MMI. Combined with the cavity effect due to the facet reflectivity, we could expect this to produce a displacement of the Fabry-Perot oscillations in one output with respect to the other.

However, systematic measurements of the transmission (similar to Fig. 3) for several fabricated samples show that the relative position of the Fabry-Perot oscillations for the two outputs actually varies from device to device. Indeed, fabrication imperfections shift the Fabry-Perot oscillations when they induce a difference of optical path of the order of $\lambda/4 \simeq 400$ nm between the two outputs. This can be due either to local imperfections (e.g. width fluctuations, impurities) or to a global difference in the geometric length of the waveguides (due e.g. to a slight non-orthogonality between the propagation and cleaving directions). 

For the study carried out in the manuscript we chose a sample that has almost identical Fabry-Perot oscillations (see Fig. 3a and b), due to a compensation of the $\pi/2$ phase difference of the MMI by fabrication imperfections. As a consequence the splitting ratio is independent of the wavelength (Fig. 3c).

For future experiments, a deterministic control over the optical paths could be achieved using the electro-optic effect \cite{Wang14}. Alternatively, an anti-reflection coating in the telecom band could be deposited on the device facets to discard any cavity effect: the transmitted power through both outputs, and thus the splitting ratio would then be independent of the wavelength, irrespective of the optical paths of the outputs.}

\section{Spectrum of the heralded single-photon source}

\begin{figure}[h]
\centering
\includegraphics[width=0.5\columnwidth]{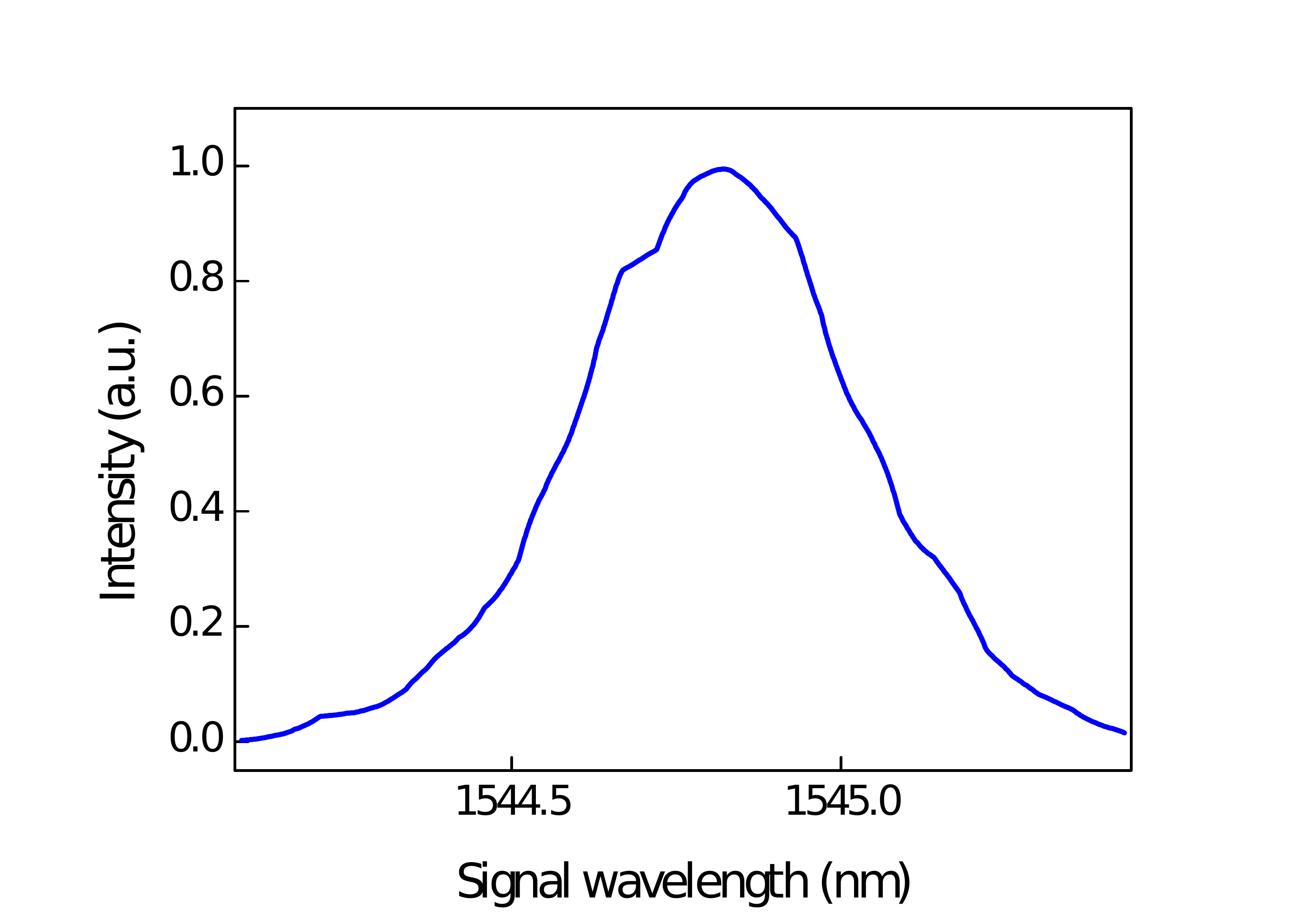}
\caption{
Experimental marginal spectrum of the heralded single photon source.
}
\label{FigSpectrum}
\end{figure}

\blue{The marginal spectrum of the heralded single-photon source can be obtained by tracing the experimental joint spectrum (Fig. 4b of the article) over the idler wavelength. The result is shown in Fig. \ref{FigSpectrum}. The measured spectral FWHM is 0.48 nm ($\simeq 60$ GHz).}

\section{Dependence of $g^{(2)}(0)$ with pump power}

In the article we measured a heralded autocorrelation function $g^{(2)}_{\rm her}(0)=0.10 \pm 0.02$ for a pump power of $P=60$ mW.
We expect that using lower pump powers should reduce the $g^{(2)}_{\rm her}(0)$ value, by further suppressing multipair emission.

Indeed, the mean photon number $\langle n \rangle$ is expected to scale linearly with $P$, while the Schmidt number $K$ is expected to remain unchanged (since it is determined by the pulse duration and spot size \cite{Caillet09,Boucher15}). We thus expect that the heralded autocorrelation function, which can be expressed as (see derivation in the section below):
\begin{equation}
g^{(2)}_{\rm her}(0)=2 \langle n \rangle \left( 1+ \frac{1}{K} \right)
\label{g2}
\end{equation}
increases linearly with $P$.

The data shown in the article demonstrate a good agreement between the value of $g^{(2)}_{\rm her}(0)$ directly measured and the one deduced from Eq. \eqref{g2}, using the experimentally determined $\langle n \rangle$ and $K$.
Building on this equivalence, we used the latter method to obtain a rapid insight into the behavior of $g^{(2)}_{\rm her}(0)$ as a function of $P$, for values of $P$ lower that the one used in the article.

\begin{figure}[h]
\centering
\includegraphics[width=0.9\columnwidth]{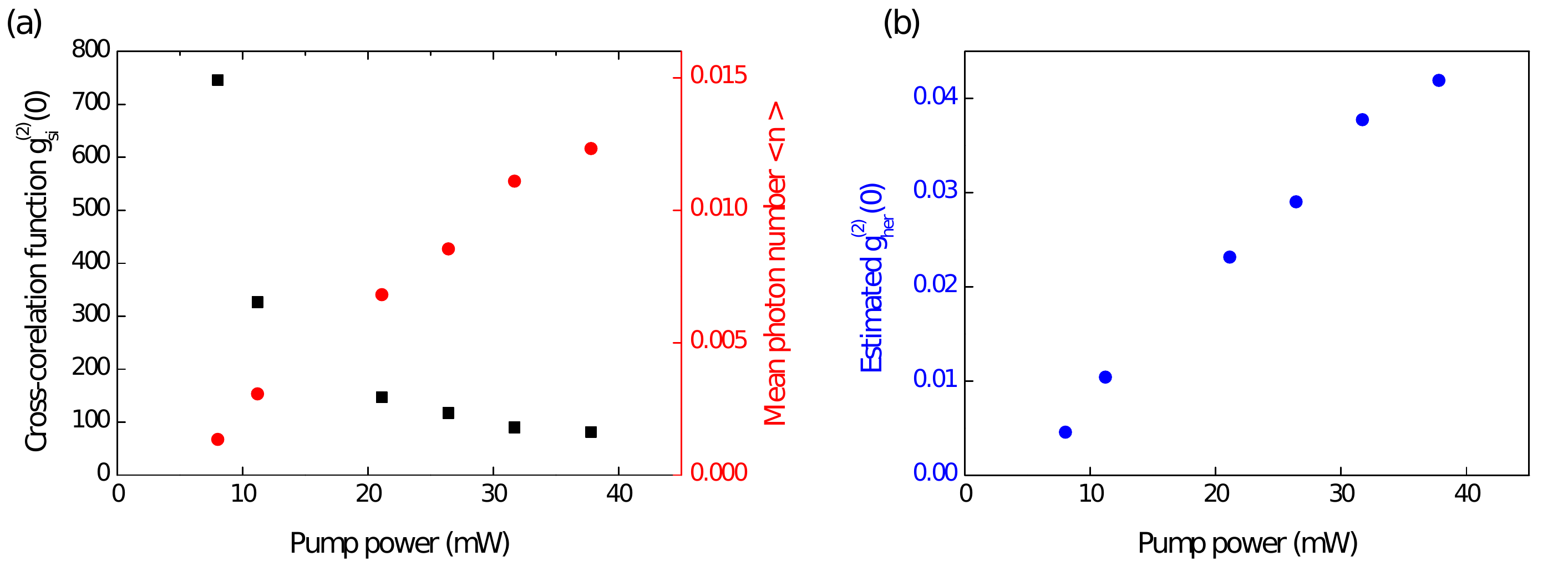}
\caption{
(a) Measured cross-correlation function (black squares) and deduced mean photon number (red points) as a function of the pump power.
(b) Estimated heralded auto-correlation function as a function of the pump power.
}
\label{Fig_g2_power}
\end{figure}

For this we measured the cross-correlation function $g^{(2)}_{s,i}(0)$ as a function of $P$ (see black squares in Fig. \ref{Fig_g2_power}a) and deduced the mean photon number using $\langle n \rangle \simeq 1/g_{s,i}^{(2)}(0)$ (red points in Fig. \ref{Fig_g2_power}a).
Finally, using the experimentally determined $K \simeq 1.4$, we used Eq. \eqref{g2} to estimate the dependence of the heralded auto-correlation function with $P$ in the low power regime, as shown in Fig. \ref{Fig_g2_power}b. The evolution is close to linear, corroborating the expectation of increased single-photon purity at smaller pump power, which could be useful for particularly demanding applications.

\section{Relationships between the heralded and unheralded autocorrelation functions}

This section presents a derivation of analytical expressions of the second-order autocorrelation function for frequency-multimode states \cite{Christ11}. These expressions, used in the article, allow to relate $g^{(2)}(0)$ to the Schmidt number $K$ and the mean photon number $\langle n \rangle$ of the states emitted by our parametric source.

\bigskip

Taking into account spectral correlations, the biphoton state generated by the SPDC process can be described by a tensor product of two-mode squeezed vacuum states \cite{Christ11}:
\begin{equation}
\ket{\Psi}=\bigotimes_{k} \sqrt{1-\lambda_{k}^{2}}\sum_{n} \lambda_{k}^{n} \ket{nn}_k \simeq \bigotimes_{i=1}^{K} \sqrt{1-\lambda^{2}}\sum_{n} \lambda^{n} \ket{nn}_i
\label{TMSV}
\end{equation}
where state vector $\ket{lj}_k$ corresponds to $j$ signal photons and $l$ idler photons in respective spectral modes denoted by $k$.
The first tensor product runs over all spectral mode $k$ arising from the Schmidt decomposition of the biphoton wavefunction; $\lambda_{k}$ is the squeezing parameter of each mode. The second equality of Eq. \eqref{TMSV} assumes that all modes have the same squeezing parameter $\lambda$; the Schmidt number $K$ is the effective number of modes needed to decompose the wavefunction.

In the low pumping regime of our experiments, $\lambda \ll 1$: we approximate the biphoton state to single and double pairs, neglecting terms with $n \ge 3$:
\begin{equation}
\ket{\Psi} \simeq \bigotimes_{i=1}^{K} \sqrt{1-\lambda^{2}} \, \left(\ket{00}_i+ \lambda \ket{11}_i + \lambda^{2} \ket{22}_i \right)
\label{TMSV5}
\end{equation}

The time-integrated second-order autocorrelation function can be expressed as:
\begin{equation}
g^{(2)}_{s,s}(0)= \frac{2P_{2}}{P^{2}_{1}}
\label{Auto3}
\end{equation}
where $P_{1}$ and $P_{2}$ are the probabilities of emitting one or two photon pairs, respectively. In our HBT configuration, the emission probability $P_{1}$ is related to the single count probability $p_{1}$ -- the probability to detect a photon from one of the two outputs of the beamsplitter -- by:
\begin{equation}
p_{1}= P_{1} \frac{\eta_{s}}{2}
\label{p1}
\end{equation}
where $\eta_{s}$ is the overall detection efficiency, accounting for both collection and APD efficiencies, and assumed to be equal for both outputs of the beamsplitter (the $1/2$ factor accounts for the splitting ratio). Similarly, the emission probability $P_{2}$ can be related to the probability $p_{2}$ to detect a coincidence at the two outputs of the beamsplitter, by:
\begin{equation}
p_{2}= P_{2} \frac{\eta_{s} ^{2}}{2}
\label{p2}
\end{equation}
Hence, the autocorrelation function can be experimentally evaluated as:
\begin{equation}
g^{(2)}_{s,s}(0)= \frac{p_{2}}{p^{2}_{1}}
\label{Auto4}
\end{equation}
as was done in the article.

\bigskip

Given the biphoton wavefunction of Eq. \eqref{TMSV5}, the probability $P_{1}$ to generate a photon pair $\ket{11}$ is, to leading order in $\lambda$:
\begin{equation}
P_{1}= K \lambda^{2}
\label{P1K}
\end{equation}
To evaluate the probability $P_{2}$ of double pair emission, we have to consider that the two photons can be either produced by two photons in the same frequency mode, with a probability $K \lambda^{4}$, or in two different modes, with a probability
$\binom{K}{2} \lambda^{2} \lambda^{2}= \frac{K(K-1)}{2} \lambda^{4}$. The total probability reads:
\begin{equation}
P_{2}= K \frac{K+1}{2} \lambda^{4}
\label{P2K}
\end{equation}
The autocorrelation function can thus be written as:
\begin{equation}
g^{(2)}_{s,s}(0)= 1+ \frac{1}{K}
\label{AutoUHK}
\end{equation}
which we used in the article to evaluate the Schmidt number of the source.

\bigskip

Let us now consider the heralded auto-correlation function:
\begin{equation}
g^{(2)}_{\rm her}(0)= \frac{2P(2\vert H)}{P(1\vert H)^{2}}
\label{AutoH}
\end{equation}
where $P(1\vert H)$ and $P(2\vert H)$ are respectively the probability to have one and two pairs produced, conditional on the detection of a heralding photon.
The approximation $P(1\vert H) \approx 1$ can be made by considering that 
most events are due to single pairs. The probability
$P(2\vert H)$ can be evaluated as the ratio $P_{s=2, H} \, / \, {p_{H}}$, where $P_{s=2, H}$ is the probability to have generated two signal photons and a heralding detection and $p_{H}$ is the probability of a heralding event. The probability of generating two pairs is $P_2$, and the probability that they give rise to a heralding count (given that our detector is not number-resolving) is $(1-(1-\eta_{H})^{2})$, where $\eta_{H}$ is the overall detection efficiency on the heralding path. Hence, $P_{s=2, H}=\frac{K(K+1)}{2}  \lambda^{4}(1-(1-\eta_{H})^{2})$. In addition, the probability of an heralding event is $K \lambda^{2} \eta_{H} $. Gathering all previous results, we obtain:
\begin{equation}
g^{(2)}_{\rm her}(0)=(K+1) \lambda^{2} \frac{1-(1-\eta_{H})^{2}}{\eta_{H}}
\label{AutoHK}
\end{equation}
In our experiments, $\eta_{H} \ll 1$, so that $1-(1-\eta_{H})^{2} \simeq 2\eta_{H}$. In this limit we obtain:
\begin{equation}
g^{(2)}_{\rm her}(0)=2 (K+1) \lambda^{2}
\end{equation}
The squeezing parameter is related to mean photon number $\langle n \rangle$ by \cite{Christ11}:
\begin{equation}
\lambda^{2}=\frac{\langle n \rangle}{\langle n \rangle + K}
\label{SqueezingParameter}
\end{equation}
In the limit $\langle n \rangle  \ll K $, we thus have:
\begin{equation}
g^{(2)}_{\rm her}(0)=2 \langle n \rangle \left( 1+ \frac{1}{K} \right)
=2\langle n \rangle \, g^{(2)}_{s,s}(0) \,
\label{AutoHK2}
\end{equation}
which is the formula used in the article to estimate the heralded autocorrelation function from the unheralded measurements.

\end{document}